\documentclass[12pt]{article}
\usepackage{epsfig}
\usepackage[centerlast,footnotesize]{caption2}
\begin{document}

\begin{center}
{\Large \bf Double Symmetry and Infinite-Component Field Theory} \\

\vspace{0.5 cm}
\renewcommand{\thefootnote}{*}
{\large L.M. Slad\footnote{E-mail: slad@theory.sinp.msu.ru}} \\

\vspace{0.4 cm}

{\it D.V. Skobeltsyn Institute of Nuclear Physics, \\
Moscow State University, Moscow 119899, Russia} 
\end{center}

\vspace{0.5 cm}

{\small Qualitative characteristics and the rigorous definition of a concept of 
the double symmetry is given. We use some double symmetry for constructing a 
theory of fields not investigated before which transform as the proper Lorentz 
group representations decomposable into infinite direct sums of 
finite-dimentional irreducible representations. All variants of the 
double-symmetric free Lagrangian of such fields are in brief described. The 
solution of the problem of changing Lagrangian mass terms due to a spontaneous 
breaking of the secondary symmetry is stated. The general properties of the 
mass spectra of fermions in the considered theory are given. It is pointed out 
a region of free parameters where the theoretical mass spectra qualitatively 
correspond to a picture typical for the parton model of hadrons.}

\vspace{0.5 cm}

\begin{center}
{\large \bf 1. Qualitative characteristics and the rigorous definition of the 
double symmetry}
\end{center}

The main objective of our work presented here is a study of the possibility 
of an alternative description of hadrons and their interactions.

It is supposed that the composite particles can have an effective description 
with the help of monolocal fields which transform as representations of the 
proper Lorentz group $L^{\uparrow}_{+}$ decomposable into infinite direct sums 
of finite-dimentional irreducible representations. We call such
infinite-component fields the ISFIR-class fields. Until recently, there was no 
study of them. It is necessary to note that the physical quantities, including
the amplitudes of processes with participation of the particles described by 
ISFIR-class fields, are expressed through infinite functional series whose 
terms are proportional to the matrix elements of finite transformations of the 
proper Lorentz group. Due to this circumstance we can expect any behaviour of 
a given physical quantity as the function of its variables, including such 
behaviour which holds in hadron physics.

The elimination of an infinite number of arbitrary parameters in the
relativistically invariant Lagrangians of the ISFIR-class fields is achieved 
due to an additional invariance of the theory. Initial (primary) and additional 
(secondary) symmetries of the theory form together the double symmetry. This 
concept can be considered as generalization for the long-known symmetry of the 
Gell-Mann--Levy $\sigma$-model [1] and supersymmetry.

In the $\sigma$-model, particles with different values of the spatial parity, 
namely, $\pi$- and $\sigma$-mesons, being a pseudoscalar and a scalar, 
accordingly, with respect to orthochronous Lorentz group $L^{\uparrow}$ are 
united in one multiplet. In oder to ensure the invariance of the theory under 
space reflection $P$ Gell-Mann and Levy have evidently declared that the
parameters of transformations, connecting between themselves the $\pi$- and 
$\sigma$-mesons, were the space pseudoscalars. The $\sigma$-model symmetry 
group $SU(2)_{L} \otimes SU(2)_{R}$ has found broad applications in the field 
theory and particle physics. However, except for Gell-Mann and Levy nobody 
spoke in obvious form about the transformation properties of these group 
parameters under the $P$-reflection.

If one reformulates the left-right symmetric model of electroweak interactions 
[2-4] in the manner of Gell-Mann--Levy, identifying some parameters of the 
local gauge group with space scalars and the others with space pseudoscalars, 
it is possible to provide the initial $P$-invariance of such model and to 
reproduce all experimentally observed consequences of the Weinberg--Salam 
model. In this case the nature of the $P$-symmetry violation appears logically 
simple and clear. If the Higgs field has both scalar and pseudoscalar 
components, and they both have nonzero vacuum expectation values whose relative
phase is not equal $\pm \pi /2$, then either left or right weak charged current
dominates depending on the relative signs of the quantities which are present 
in gauge transformations. It means that, as a result of spontaneous breaking of
the local symmetry, the physical vacuum does not possess a certain $P$-parity. 
The fields of all intermediate bosons constitute a superposition of polar and 
axial 4-vectors, these vectors having equal weights in the fields of 
$W$-bosons. Transformation properties of the gauge bosons with respect to the 
orthochronous Lorentz group have the same powerful significance as the form of 
weak currents. However, they yield no experimental detection in contrast to the
interaction vertices.

In Ref. [5], the grain of new group theoretical approach sown by Gell-Mann and 
Levy has received its generalization in the concept of double symmetry 
consisting of the primary and secondary symmetries.

The secondary symmetry has three essential features. First, the parameters of
the transformations generating a global or local secondary-symmetry group 
$H_{T}$ belong to the given $T$-representation space of the global group of the
primary symmetry $G$. Second, the $H_{T}$-group transformations operate on 
the vectors of some $S$-representation space of the group $G$ and do not take 
out them from this space, i.e. the secondary symmetry does not violate the 
primary one. Third, the primary- and secondary-symmetry groups $G$ and $H_{T}$ 
have no common elements except for the unity, and even in that case when they 
are isomorphic.

We restrict ourselves with the consideration of the global symmetries. The 
above-listed features find their realization in the following rigorous 
definition.

\vspace{0.2cm}

{\bf Definition.} {\it Let the group ${\cal G}_{T}$ contain the subgroup $G$ 
and the invariant subgroup $H_{T}$, with $G \neq {\cal G}_{T}$, 
$H_{T} \neq {\cal G}_{T}$, and be their semidirect production 
${\cal G}_{T} = H_{T} \circ G$. If any element $h_{T} \in H_{T}$ can be written
in the form
\begin{equation}
h_{T} = h(\theta_{1})  h(\theta_{2}) \ldots h(\theta_{n}), \hspace{0.5 cm}
n \in \{1, 2, \ldots \}, 
\end{equation}
where $\theta_{i}$ $(i = 1, 2, \ldots, n)$ is some vector of the representation 
space of $T$ of the group $G$, and if for any element $g \in G$
\begin{equation}
g h(\theta) g^{-1} = h(T(g)\theta ),
\end{equation}
then the group $G$  will be called the primary symmetry group, and the groups 
$H_{T}$ and ${\cal G}_{T}$ will be respectively called the secondary and 
double symmetry groups generated by the representation $T$ of the group $G$.}

\vspace{0.2cm}

Constructing the double-symmetric field theories is based on the following 
consequence of the secondary-symmetry definition.

\vspace{0.2cm}

{\bf Corollary.} {\it Let $\Psi(x)$ be any field vector in the 
$S$-representation space, and $\theta =  \{\theta_{a}\}$ be a vector in the 
$T$-representation space of the group $G$. Then the transformation
\begin{equation}
\Psi '(x) = \exp (-iD^{a}\theta_{a})\Psi (x)
\end{equation}
will be a secondary-symmetry transformation iff the condition
\begin{equation}
D^{a} = S^{-1}(g) D^{b} S(g) [T(g)]_{b}{}^{a}
\end{equation}
is fulfilled, i.e. the operators $D^{a}$ are the $T$-operators (transform as
the representation $T$) of the group $G$.}

\vspace{0.2cm}

In our construction [6,7] of the theory of infinite-component fields of the 
ISFIR-class the secondary symmetry is generated by transformations (3), with 
the index $\mu$ instead of the index $a$, in which the parameters 
$\theta_{\mu}$ are the components of the polar or axial 4-vectors of the 
orthochronous Lorentz group, and the operators $D^{\mu}$ have the matrix 
realization. We postulate no closing of the algebra of the operators $D^{\mu}$.
It arises automatically in each variant of the considered theory due to those 
restrictions which are imposed by the requirement that the corresponding 
Lagrangians possess the double invariance.

\begin{center}
{\large \bf 2. Double-symmetric Lagrangians of the free ISFIR-class fields}
\end{center}

The general structure of relativistically invariant free Lagrangians of the 
form
\begin{equation}
{\cal L}_{0} = \frac{i}{2}[(\Psi, \Gamma^{\mu} \partial_{\mu} \Psi ) - 
(\partial_{\mu} \Psi, \Gamma^{\mu} \Psi )] - (\Psi, R \Psi )
\end{equation}
was found and described by Gelfand and Yaglom [8,9]. Let us remind that, 
according to Gelfand and Yaglom, the finite-dimentional irreducible 
representation $\tau$ of the proper Lorenz group is defined by such a pair of 
numbers $(l_{0},l_{1})$ that $2l_{0}$ and $2l_{1}$ are integers of the same 
parity and $|l_{1}| > |l_{0}|$. The canonical basis vectors  of this 
representation space are related to the subgroup $SO(3)$ and are denoted by 
$\xi_{(l_{0},l_{1})lm}$ where $l$ is a spin, $m$ is its projection onto the 
third axis, $m=-l,-l+1,\ldots, l$, and $l=|l_{0}|,|l_{0}|+1,\ldots, |l_{1}|-1$.
The pairs $(l_{0},l_{1})$ and $(-l_{0},-l_{1})$ define the same 
representations, $(l_{0},l_{1}) \sim (-l_{0},-l_{1})$.

The requirement that the Lagrangian (5) is invariant under the group 
$L^{\uparrow}_{+}$ gives for the operator $\Gamma^{\mu}$ a condition which is 
equivalent to the relation (4) (at $g \in L^{\uparrow}_{+}$) for the operator 
$D^{\mu}$ from the secondary-symmetry transformation. It has the consequences 
expressed through the following equalities
\begin{equation}
\Gamma^{i} = [I^{i0}, \Gamma^{0}] \hspace{0.5 cm} (i=1,2,3),
\end{equation}
$$\Gamma^{0} \xi_{(l_{0}, l_{1}) lm} = c(l_{0}+1,l_{1};l_{0},l_{1})
\sqrt{(l+l_{0}+1)(l-l_{0})} \xi_{(l_{0}+1, l_{1})} +$$
$$+ c(l_{0}-1,l_{1};l_{0},l_{1}) \sqrt{(l+l_{0})(l-l_{0}+1)} 
\xi_{(l_{0}-1, l_{1}) lm} +$$
$$+ c(l_{0},l_{1}+1;l_{0},l_{1}) \sqrt{(l+l_{1}+1)(l-l_{1})}
\xi_{(l_{0}, l_{1}+1) lm} +$$
\begin{equation}
+ c(l_{0},l_{1}-1;l_{0},l_{1}) \sqrt{(l+l_{1})(l-l_{1}+1)}
\xi_{(l_{0}, l_{1}-1) lm},
\end{equation}
where $I^{i0}$ are the infinitesimal operators of the proper Lorentz group, and
$c(l'_{0},l'_{1};l_{0},l_{1})$ $\equiv c_{\tau' \tau}$ are arbitrary 
parameters. The same equality are valid for the operator $D^{\mu}$ with 
arbitrary quantities $d_{\tau' \tau}$.

Infinite number of arbitrary parameters $c_{\tau' \tau}$ in the 
relativistically invariant Lagran- gians of the free ISFIR-class fields was the 
serious reason for that until recently there were no research of such field 
theory. 

Involved into our consideration are the $L^{\uparrow}_{+}$-group 
representations $S$, in whose decomposition the multiplicity of each 
finite-dimentional irreducible representation does not exceed unity.

The Lagrangian (5) (at $R \neq 0 $) will be invariant under the 
secondary-symmetry transformations (3) generated by the polar or axial 4-vector
of the group $L^{\uparrow}$, if the condition 
\begin{equation}
[\Gamma^{\mu}, D^{\nu}] = 0
\end{equation}
is satisfied. This condition is reduced to an algebraic system of the equations
with respect to quantities $c_{\tau' \tau}$ and $d_{\tau' \tau}$. 

As the analysis shows this system of the equations has nontrivial solutions 
only for the quite certain countable set of the $L^{\uparrow}_{+}$-group
representations $S$. The fermionic field can be described either by any of the
representations
\begin{equation}
S^{k_{1}} = \sum^{+\infty}_{n_{1}=0} \sum^{k_{1}-3/2}_{n_{0}=-k_{1}+1/2}
\oplus (\frac{1}{2}+n_{0}, k_{1}+n_{1}),
\end{equation} 
where $k_{1} \geq 3/2$, or by the representation $S^{F}$ which contains all
finite-dimentional irreducible half-integer spin representations of the group
$L^{\uparrow}_{+}$. The bosonic field can correspond either to any of the
representations
\begin{equation}
S^{k_{1}} = \sum^{+\infty}_{n_{1}=0} \sum^{k_{1}-1}_{n_{0}=-k_{1}+1}
\oplus (n_{0}, k_{1}+n_{1}),
\end{equation}
where $k_{1} \geq 1$, or to the representation $S^{B}$ which contains all
finite-dimentional irreducible integer spin representations of the group
$L^{\uparrow}_{+}$.

Three variants of the theory are assigned to each of representations (9). In 
one of them the secondary symmetry is generated by a polar 4-vector of the 
group $L^{\uparrow}$, the operators $D^{\mu}$ commute with each other
\begin{equation}
[D^{\mu},D^{\nu}] = 0,
\end{equation} 
i.e. the group $H_{T}$ is the Abelian one, and the quantities $c_{\tau' \tau}$
are given by equalities
$$c(l_{0}+1,l_{1};l_{0},l_{1}) = c(l_{0},l_{1};l_{0}+1,l_{1}) =$$
\begin{equation}
= c_{0} D(l_{1}) \sqrt{\frac{(k_{1}-l_{0}-1)(k_{1}+l_{0})}
{(l_{1}-l_{0})(l_{1}-l_{0}-1) (l_{1}+l_{0}) (l_{1}+l_{0}+1)}},
\end{equation}
$$c(l_{0},l_{1}+1;l_{0},l_{1}) = c(l_{0},l_{1};l_{0},l_{1}+1) =$$
\begin{equation}
= c_{0} D(l_{0}) \sqrt{\frac{(k_{1}-l_{1}-1)(k_{1}+l_{1})}
{(l_{1}-l_{0})(l_{1}-l_{0}+1) (l_{1}+l_{0}) (l_{1}+l_{0}+1)}},
\end{equation}
where $D(j) = 1$, and $c_{0}$ is a real constant. In the second and third 
variants of the theory the secondary symmetry is generated by an axial 4-vector
of the group $L^{\uparrow}$. In the second variant, the formulas 
(9)--(11), where $D(j) = (-1)^{j-1/2} j$, are valid. In the third variant it is
held the inequality
\begin{equation} 
[G^{\mu\nu}, G^{\rho\sigma}] \neq 0,
\end{equation}
in which the antisymmetric operator $G^{\mu\nu}$ is involved. This operator is
determined as follows 
\begin{equation}
G^{\mu\nu} = [D^{\mu}, D^{\nu}].
\end{equation}

Four variants of the theory at $k_{1} \geq 2$ and two variants at $k_{1} = 1$ 
are assigned to each of representations (10). All variants divides into pairs 
in which the secondary symmetry is generated either by a polar, or by an axial 
4-vector of the group $L^{\uparrow}$. In one pair of the variants 
$k_{1} \geq 1$ and the equality (11) is valid, and in other pair $k_{1} \geq 2$
and the inequality (14) holds.

The representation $S^{F}$ is assigned by one variant of the theory, and
the representation $S^{B}$ is by two variants. In these cases the operators 
$G^{\mu\nu}$ are nonzero and commute with each other, and the 
secondary-symmetry transformations of fermionic (bosonic) fields is generated 
by the axial (polar or axial) 4-vector of the group $L^{\uparrow}$.

In all variants of the double-symmetric theory the operator $R$, specifying 
the mass term of Lagrangian (5), is a multiple of the identity operator.

\begin{center}
{\large \bf 3. A spontaneous secondary-symmetry breaking}
\end{center}

In complete accordance with the known Coleman--Mandula theorem [10] relating to
consequences of the Lorentz-group extension, the mass spectrum corresponding to
any of the variants of the infinite-component field theory with the double 
symmetry is infinitely degenerate with respect to spin. This degeneracy forces 
us to suppose that the secondary symmetry is spontaneously broken, namely, that
the scalar (with respect to the group $L^{\uparrow}$) components of one or 
several bosonic infinite-component ISFIR-class fields have nonzero vacuum 
expectation values $\lambda^{i}$. A spontaneous secondary-symmetry breaking 
introduces only one correction into the free-field theory with the initial 
double symmetry which consists in changing the operator $R$ in the Lagrangian 
(5). So, in the fermionic-field Lagrangian, this operator takes the form 
\begin{equation}
R = \kappa E + \lambda^{i} Q^{(0,1)00}_{i},
\end{equation}
where $\kappa$ is a constant. The $P$-even operators $Q^{(0,1)00}_{i}$ come to 
relation (16) from the interaction Lagrangian
\begin{equation}
{\cal L}_{\rm int} =   
(\psi (x), Q^{\tau lm}_{i'} \varphi_{\tau lm}^{i'} (x) \psi (x)),
\end{equation}
where $\psi (x)$ is the fermionic field, $\varphi_{\tau lm}^{i'} (x)$ is the
bosonic-field component, and $Q^{\tau lm}_{i}$ are the matrix operators.

The Lagrangian (17) containing nonzero operators $Q^{(0,1)00}_{i}$ possesses 
the considered double symmetry only in two cases denoted by ${\cal A}$ and 
${\cal B}$. In each of them the secondary-symmetry group is the 
four-parametrical Abelian group, the fermionic field transforms as any of the 
representations $S^{k_{1}}$ (9), and bosonic fields transform as the 
representation $S^{1}$ (10). In the case $ {\cal A} $, the secondary symmetry 
is generated by the polar 4-vector of group $L^{\uparrow}$, and $D(j) = 1$. In 
the case ${\cal B}$, the secondary symmetry is generated by the axial 4-vector 
of group $L^{\uparrow}$ and $D(j) = (-1)^{j-1/2} j$. In both cases the 
relations
\begin{equation}
R \xi_{(l_{0},l_{1})lm} = r(l_{0},l_{1}) \xi_{(l_{0},l_{1})lm} =
[ \kappa + \lambda^{i} q_{i}(l_{0},l_{1}) ] \xi_{(l_{0},l_{1})lm},
\end{equation}
$$(k_{1}-l_{0}-1)(k_{1}+l_{0})q_{i}(l_{0}+1,l_{1})
+ (k_{1}-l_{0})(k_{1}+l_{0}-1)q_{i}(l_{0}-1,l_{1}) -$$
$$- (k_{1}-l_{1}-1)(k_{1}+l_{1})q_{i}(l_{0},l_{1}+1)
- (k_{1}-l_{1})(k_{1}+l_{1}-1)q_{i}(l_{0},l_{1}-1) =$$
\begin{equation}
= z_{i}(l_{1}-l_{0})(l_{1}+l_{0})q_{i}(l_{0},l_{1})
\end{equation}
are valid for all irreducible representations $(l_{0},l_{1}) \in S^{k_{1}}$. 
The parameter $z_{i}$ is expressed through the ratio of normalization constants 
of the operators $D^{\mu}$ in the secondary-symmetry transformations (3) of the
bosonic and fermionic fields. It can have any values. If $z_{i} = 2$, then the 
bosonic-field transformation (3) is trivial ($D^{\mu} = 0$), i.e. it is not 
involved in the secondary symmetry.

In what follows we will deal only with fermionic fields which transform
according to the "lowest" of the representations $S^{k_{1}}$ (9) of the proper 
Lorentz group, namely, according to the representation $S^{3/2}$. In this case 
the equation (19) has a unique solution which can be written in the form
\begin{equation}
q_{i}\left( -\frac{1}{2},l_{1}\right) = q_{i}\left( \frac{1}{2},l_{1}\right)  = 
2q_{i0}\frac{u_{i}^{N}(u_{i}N+N+1)-w_{i}^{N}(w_{i}N+N+1)}{N(N+1)(u_{i}-w_{i})
(2+u_{i}+w_{i})},
\end{equation}
where $N = l_{1}-1/2$, $u_{i}=(z_{i}+\sqrt{z_{i}^{2}-4})/2$ and 
$w_{i}= (z_{i}-\sqrt{z_{i}^{2}-4})/2$.

\begin{center}
{\large \bf 4. Properties of the mass spectra in the ISFIR-class field theory} 
\end{center}

We have all necessary for solving of the first physical problem which concerns 
the general properties of the mass spectra in the double-symmetric theory of 
the ISFIR-class fermionic fields.

In the rest system of a particle with the mass $M$ the Gelfand-Yaglom 
equation corresponding to the Lagrangian (5) has the form
\begin{equation}
(M \Gamma^{0} - R) \psi_{M0} = 0.
\end{equation}
Its solutions must satisfy the selection condition at which the amplitudes of 
various processes are finite. We formulate such condition of finiteness of the 
amplitudes as the relation
\begin{equation}
|(\psi_{M0}, R \psi_{Mp})| < +\infty
\end{equation}
or 
\begin{equation}
(\psi_{M'0}, R \psi_{Mp}) = a(p) \delta (M'-M),
\end{equation}
where $\psi_{Mp}$ is the particle state vector with the momentum $p$ directed 
along the third axis, and $a(p)$ is a nonzero number.

In the formula (22) or (23) the bilinear form is expressed through the infinite
series whose terms are linear with respect to the components of vector 
$\psi_{M0}$, to the components of other vector $\psi_{M0}$ or $\psi_{M'0}$, and
to the matrix elements of finite transformations of the proper Lorentz group. 
These matrix elements for irreducible representation $(l_{0},l_{1})$ behave as 
$T_{0} \exp (\alpha l_{1})/l_{1}$ at $l_{1} \rightarrow +\infty$ where
$\tanh \alpha =p/\sqrt{M^{2}c^{2}+p^{2}}$, and the quantity $T_{0}$ does not 
depend on the number $l_{1}$ [11].

In the space of the $L^{\uparrow}_{+}$-group representation $S^{3/2}$ the
equation (21) for the field vector components 
$(\psi_{M0})_{(\pm 1/2,l_{1})lm}\equiv\chi_{lm}(l_{1})$ with the spatial parity
 $(-1)^{l-1/2}$ takes the form
$$D\left( \frac{1}{2} \right) \frac{\sqrt{(l_{1}-l)(l_{1}+l+1)}}{2l_{1}+1} 
\chi_{lm} (l_{1}+1)
+D\left( \frac{1}{2} \right) \frac{\sqrt{(l_{1}-l-1)(l_{1}+l)}}{2l_{1}-1} 
\chi_{lm} (l_{1}-1 ) -$$
\begin{equation}
- \left[ \frac{D(l_{1})(2l+1)}{4l_{1}^{2}-1}
- \frac{1}{2Mc_{0}} r(l_{1}) \right] \chi_{lm} (l_{1}) = 0,
\end{equation}
where $l_{1} \geq l$, $r(l_{1}) \equiv r(\pm 1/2,l_{1})$. At the replacement of
the value $M$ by $-M$ the $P$-parity of solution of this equation changes by 
the opposite one.

In a situation when the secondary symmetry of the theory is broken 
spontaneously we failed either to find solutions of the equation (24) in the 
form of elementary or special functions, or to find analytical formulas for 
mass spectra of the theory. However, we are able to make a number of 
conclusions concerning the mass spectra on the basis of asymptotic behaviour of
some quantities. Being based on these conclusions it is possible to find any 
number of the lower values of the mass $M$ with the help of numerical methods.

We have analysed the mass spectrum characteristics in all details in both cases 
${\cal A}$ and ${\cal B}$ of the considered theory provided that the 
spontaneous breaking of the secondary symmetry is caused by one ISFIR-class 
bosonic field, and $\kappa = 0$ in formulas (16) and (18).

In the region of the parameter values $z_{1} \in (-2,2)$ the solutions of 
equation (24) can not satisfy the condition of finiteness of the amplitudes 
(22) or (23) at any values of the quantity $M$, i.e. the mass spectrum is 
empty.

In the region of the parameter values $z_{1}\in (-\infty,-2] \cup (2,+\infty)$
the set of all masses of the theory is restricted from below by some positive 
number. At $l_{1} \rightarrow +\infty$ the quantities $r(l_{1})$ and 
$\chi_{lm}(l_{1})$ have the following asymptotic behaviour
\begin{equation}
r(l_{1}) = r_{0} \frac{v^{l_{1}+\frac{1}{2}}}{l_{1}+\frac{1}{2}} 
(1+{\cal O}(l_{1}^{-1})),
\end{equation}
\begin{equation}
\chi_{lm}(l_{1}) = A_{0} G(l_{1}) (1+{\cal O}(l_{1}^{-1})) 
+ B_{0} G^{-1}(l_{1}) (1+{\cal O}(l_{1}^{-1})),
\end{equation}
where
\begin{equation}
G(l_{1}) = \frac{x^{l_{1}-\frac{1}{2}}v^{\frac{4l_{1}^{2}-1}{8}}}
{(l_{1}-\frac{1}{2})!},
\hspace{0.5 cm} x = -\frac{r_{0}}{Mc_{0}D(\frac{1}{2})}, \hspace{0.5 cm}
v = \left\{ 
\begin{array}{ll}
w_{1}, & {\mbox{\rm if}} \hspace{0.3cm} z_{1} \in (-\infty,-2] \\
u_{1}, & {\mbox{\rm if}} \hspace{0.3cm} z_{1} \in (2, +\infty).
\end{array} \right. 
\end{equation}

If the quantity $A_{0}$ is not equal to zero at some value of $M$, then neither
condition (22), nor condition (23) cannot be fulfilled. If $A_{0}=0$ at some 
value of $M$, then the condition (22) is fulfilled. Hence, for all 
$z$-parameter values from the region $z \in (-\infty, -2)$, the mass spectrum 
is discrete, if it is not empty. As the calculations show in the whole 
parameter region $z > 2$ each value of spin and parity corresponds to an 
accountable set of the masses rising to infinity. The lowest values of mass 
levels, corresponding to a definite spin, grow with a spin. Thus, in the 
qualitative sense the theoretical mass spectra correspond to a picture typical 
for the quark-gluon model of hadrons.

We compare the theoretical mass spectrum with nucleon resonance levels provided 
that the spontaneous breaking of the secondary symmetry is caused by two 
ISFIR-class bosonic fields, and $\kappa = 0$ in formulas (16) and (18). In this
situation we pay attention to the following property of the particle state 
vectors in the considered theory.

In the reference system where the ground fermion (boson) of the theory 
possesses a nonzero velocity, its state vector has nonzero components with all 
half-integer (integer) spins, irrespective to what spin corresponds to this
particle in its rest system. This results in that the amplitude of the 
resonance decay into ground particles will contain any partial wave in its 
decomposition. Certainly, only those of them may be observable in experiments
which have an appreciable weight. Thus, one resonance can be shown in 
experiments as a group of several resonances with the same mass and 
differing only in their spin.

\begin{center}
\vspace{-0.8cm}
\begin{figure}[ht]
\includegraphics[width=7.85cm]{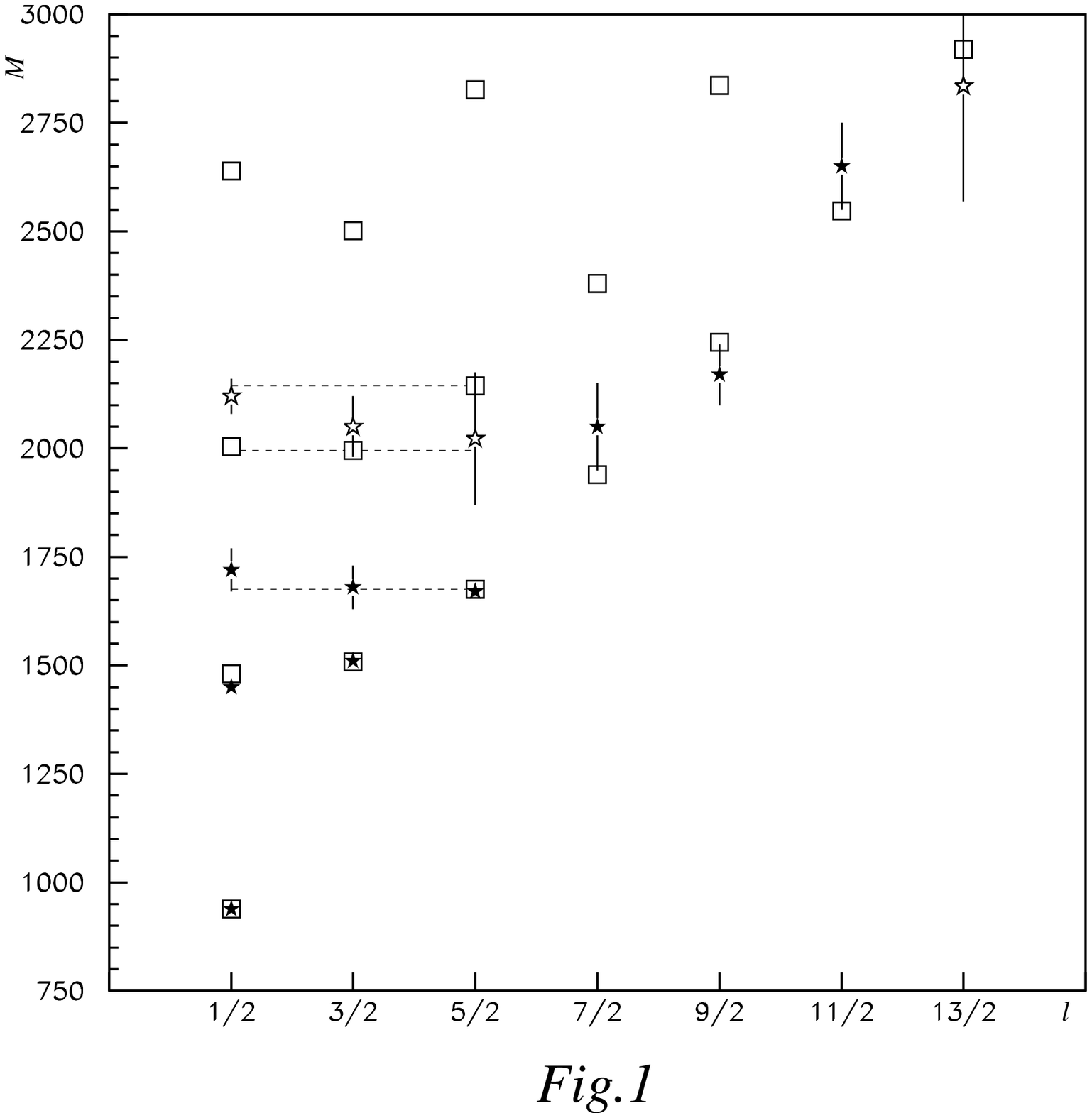}
\includegraphics[width=7.85cm]{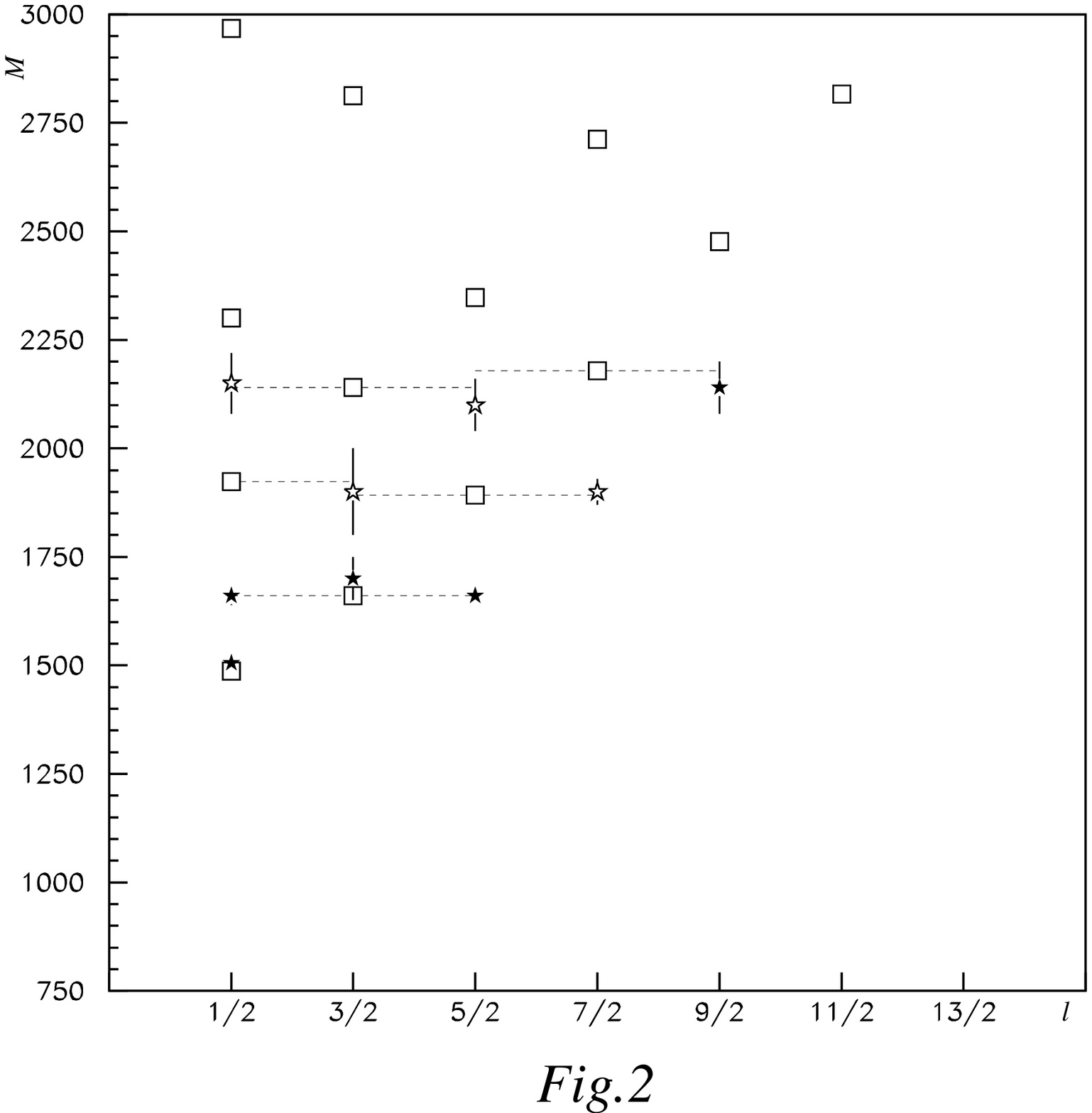}
\end{figure}
\end{center}

The results given in Figs. 1 and 2 correspond to the case ${\cal A}$ of the 
considered theory at the following choice of parameters:
$\lambda^{1}q_{10}/c_{0} = -939/2.4686$ MeV, $z_{1} = 2.036$, $z_{2} = 0.14$, 
$\lambda^{2}q_{20}/\lambda^{1}q_{10} = -0.6724$. The levels with the partial
parities $(-1)^{l-1/2}$ and $(-1)^{l+1/2}$ are represented in Figs. 1 and 2
correspondingly. The theoretical masses are depicted by the square markers, and
the experimental masses (the pole positions) are done by the dark- and 
light-star markers for, accordingly, well and badly established nucleon 
resonances [12]. The Breit-Wigner mass is taken for the resonance $N(1440)$ 
because two poles are found around 1440 MeV, $1370-114i$ and $1360-120i$, whose
parameters are much different from the conventional ones $M = 1470$ MeV and 
$\Gamma = 545$ MeV [13]. The first approximation of the theory to experiments 
can be estimated as satisfactory.

\vspace{0.3 cm}

{\bf Acknowledgements.} I am very grateful to S.P.~Baranov, E.E.~Boos, 
V.I.~Fushchych, A.U.~Klimyk, A.A.~Komar, V.I.~Savrin, I.P.~Volobuev and 
N.P.~Yudin for useful discussions of my work holding in different time.


\begin{thebibliography}{99}

\bibitem{1}
   M.Gell-Mann and M.Levy, {\it Nuovo Cimento} {\bf 16} (1960) 705.
\bibitem{2}
   J.C.Pati and A.Salam, {\it Phys.Rev.} {\bf D10} (1974) 275.
\bibitem{3}
   R.N.Mohapatra and J.C.Pati, {\it Phys.Rev.} {\bf D11} (1975) 566 and 2558.
\bibitem{4}
   G.Senjanovic and R.N.Mohapatra, {\it Phys.Rev.} {\bf D12} (1975) 1502.
\bibitem{5}
   L.M.Slad, {\it Mod.Phys.Lett.} {\bf A15} (2000) 379 (hep-th/0003107).
\bibitem{6}
   L.M.Slad, {\it Theor.Math.Phys.} {\bf 129} (2001) 1369 (hep-th/0111140). 
\bibitem{7}
   L.M.Slad, {\it Theor.Math.Phys.}, {\bf 133} (2002) 1363 (hep-th/0210120).
\bibitem{8}
   I.M.Gelfand and A.M.Yaglom, {\it Zh.Eksp.Teor.Fiz.} {\bf 18} (1948) 703.   
\bibitem{9}
   I.M.Gelfand, R.A.Minlos, and Z.Ya.Shapiro, {\it Representations of the 
   rotation and Lorenz group and their applications} (The Macmillan Company, 
   New York, 1963).
\bibitem{10}
   S.Coleman and J.Mandula, {\it Phys.Rev.} {\bf 159} (1967) 1251.   
\bibitem{11}
   S.Str\"{o}m, {\it Arkiv f.Fysik} {\bf 29} (1965) 467.
\bibitem{12}
   Particle Data Group, K.Hagiwara et al., {\it Phys.Rev.} {\bf D66} (2002) 
   010001.    
\bibitem{13}
   R.E.Cutkosky and S.Wang, {\it Phys.Rev.} {\bf D42} (1990) 235.
   
\end{thebibliography}
\end{document}